# On the distance dependence of electron transfer through molecular bridges and wires

Spiros Skourtis
Department of Physics, University of Cyprus, 1678 Nicosia, Cyprus

and

Abraham Nitzan
School of Chemistry, Tel Aviv University, Tel Aviv, 69978, Israel

## Abstract

The dependence of electron transfer rates and yields in bridged molecular systems on the bridge length, and the dependence of the zero-bias conduction of molecular wires on wire length are discussed. Both phenomena are controlled by tunneling across the molecular bridge and are consequently expected to show exponential decrease with bridge length that is indeed often observed. Deviations from this exponential dependence for long bridges, in particular a crossover to a very weak dependence on bridge length were recently observed experimentally and discussed theoretically in terms of thermal relaxation and dephasing on the bridge. Here we discuss two other factors that potentially affect the bridge length dependence of these phenomena. First, in experiments initiated by an initial preparation of a non-stationary "donor" state the initial energy is not well defined. A small contribution from initially populated eigenstates that are of mostly bridge-level character may dominate transmission for long bridges, resulting in weak or no bridge-length dependence. Secondly, in steady state experiments the distribution of initial states (for example the Fermi distribution at the electrodes in conduction experiments) will cause deviations from exponential dependence on wire length because even a small population in higher energy states will dominate the transmission through long wires. For the first scenario we show that the crossover behavior observed for electron transfer in DNA between G and GGG species separated by AT chains can be largely reproduced just by initial state effects.

## 1. Introduction

The distance dependence of electron transfer rates and yields is obviously an important attribute of the process.[1] The tunneling nature of this transfer is manifested in a characteristic exponentially decreasing behavior with increasing bridge length. Figure 1 shows a well-known simple model for this phenomenon. In Figure 1a the electron transfer takes place between donor (D) and acceptor (A) species through a molecular bridge B represented by a set of *N* consecutive levels with nearest-neighbor coupling. In Figure 1b the donor and acceptor are replaced by two metal electrodes, represented by dense manifolds of (quasi-free) electron states. Both models are characterized by the bridge length *N*, the coupling *V* of the first and last bridge levels to the donor and acceptor (or the leads) states, respectively, the bridge inter-level coupling $V_B$ and the electronic energy gap $\Delta E_B$ between the bridge energy $\varepsilon_B$ and the injection energy (i.e. the donor energy in the electron-transfer system and the Fermi energy in the conduction case). For simplicity we take same nearest neighbor couplings in the bridge and same couplings between the bridge and the donor/acceptor species, and also take the same donor and acceptor energies, $\varepsilon_D = \varepsilon_A$ and a single bridge energy. The conclusions we reach below do not depend on these simplifications. The Hamiltonian of the DBA system is

$$H_{DBA} = \varepsilon_0 |0\rangle\langle 0| + \varepsilon_{N+1} |N+1\rangle\langle N+1| + \sum_{l=1}^{N} \varepsilon_l |l\rangle\langle l| \\ + (V|0\rangle\langle 1| + hc) + (V|N+1\rangle\langle N| + hc) + \sum_{l=1}^{N-1} (V_B |l\rangle\langle l+1| + hc) \quad (1)$$

We use interchangeably the notations D and A or 0 and *N*+1, respectively. For transition between two metals The super-exchange limit is often referred to the case where $|\Delta E_B| \gg |V_B|$. In this case both the rate in the model of Fig. 1a and the zero-bias conduction in Fig. 1b are approximately proportional to $(V_B/\Delta E_B)^{2N}$, implying length dependence

$$rate, yield, current \sim \exp(-\beta N) \; ; \quad \beta = 2\ln(\Delta E_B / V_B) . \quad (2)$$

The parameter $\beta$ depends on the particular bridge molecule used. For many molecules it lies in the range 0.5 - 2.0.[2]

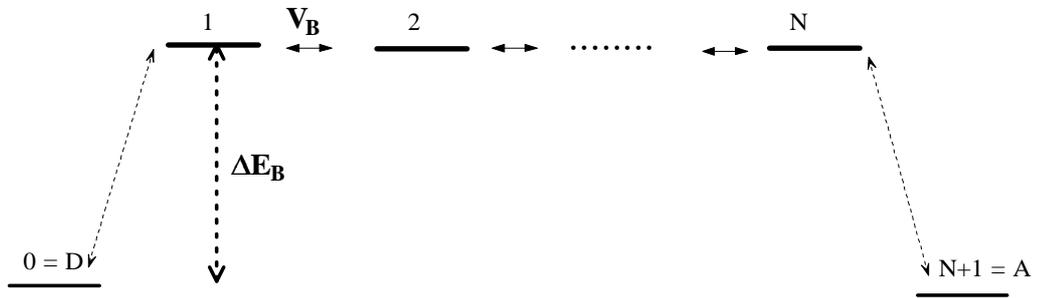

Fig. 1a

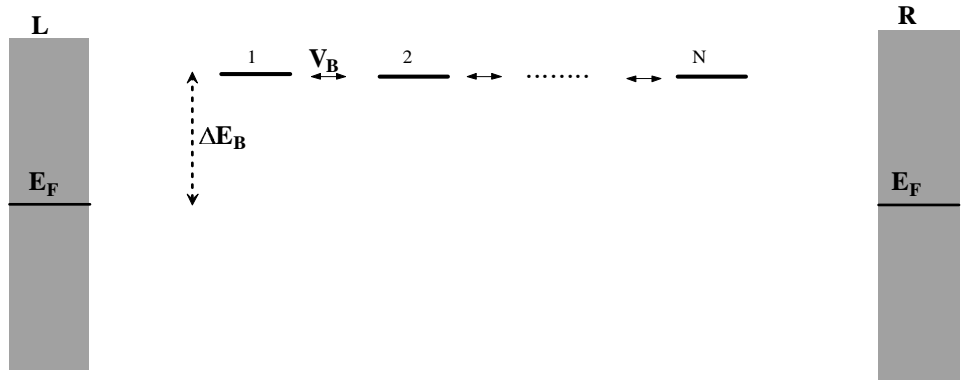

Fig. 1b

Fig. 1. Energy levels diagrams for the model considered in the present discussion: (1a) electron transfer between a donor (D) and acceptor (A) states. (1b) Electron transmission between two electrodes (continuous manifolds of states represented by the gray areas. See text for further details.

Such exponential length dependence qualifies a molecular wire as an insulator. In contrast, coherent transfer at resonance ($\Delta E_B = 0$) does not depend on *N*, while classical conducting behavior shows an Ohmic length dependence of the conduction *g*, $g \sim N^{-1}$. Classical conducting behavior occurs when dephasing interactions dominate the resonant transfer, transforming a ballistic motion into a hopping, essentially diffusive, transfer.[1]

In addition to these coherent and incoherent transfer limits, interesting intermediate cases have been discussed theoretically and demonstrated experimentally. These cases show a crossover from the exponential $N$ dependence that characterizes off-resonance coherent transfer, to a behavior described by $(A+BN)^{-1}$ with A and B constants. In phenomenological approaches this behavior is accounted for by a superposition of two processes, coherent tunneling that dominates the transfer for small $N$ and drops exponentially as $N$ increases and activation onto the bridge followed by incoherent hops along it.[3,4] The constant A is related to the thermal activation time, and for intermediate values of $N$ it may happen that $A \gg BN$ and that the transfer rate or yield beyond the crossover from the exponential behavior may appear practically independent of the bridge length. (A residual exponential drop may be imposed on this kinetic regime if the electron is exposed to additional loss processes on the bridge (for example electron capture by a surrounding solvent, e.g. water). In the limit of large $N$ the transfer (or conduction) assumes the Ohmic, $N^{-1}$ behavior. Table 1 summarizes these processes.

**Table 1  Bridge length dependence of the transmission rate**

| Physical Process | Bridge length ($N$) dependence | Comments |
|---|---|---|
| **Super exchange (short bridge, $\Delta E_B \gg V_B$)** | $e^{-\beta N}$ | $\beta = 2\ln(\Delta E_B / V_B)$ |
| **Steady state hopping (long bridge)** | $N^{-\delta}$ | $\delta$ depends on boundary conditions. $\delta = 1$ for directional hopping between source and sink |
| **Intermediate range (intermediate length bridge)** | $(A+BN)^{-1} e^{-\Delta E_B/k_B T}$ | for directional hopping |
| **Steady state hopping + competing loss at every bridge site** | $e^{-\alpha N}$ | $\alpha$ is related to electron loss rate |

More rigorous treatments of electron transfer[5,6] show that the different processes displayed in Table 1 do not contribute to the overall transmission in a simple additive way, however the qualitative behavior is the same and the insight obtained from the qualitative picture remains useful. These theoretical treatments incorporate into the quantum dynamics implied by the Hamiltonian (1) additional interactions with the thermal environment. This results in activation and dephasing transitions that, in the limit of strong thermal interactions, leads to the incoherent hopping dynamics.

Recent experimental observations of the crossover behavior described above lend support for this thermal mechanism. However, here we point out that crossover from exponential decrease with bridge length to a weak length dependence may arise also from other physical origins. In the present paper we examine two such possibilities:

(a) For processes that are initiated by a sudden preparation of the initial donor state the actual initial energy is not well defined because this state is not an eigenstate of the DBA Hamiltonian. Some (small) population must be placed on eigenstates of the DBA systems that are delocalized on the bridge. The probability that this population is detected on the other side of the bridge does not depend on the bridge length. For long enough bridge this population can dominate the observed transfer.

(b) For processes in which thermal relaxation on the donor and acceptor sites is fast while that on the bridge can be disregarded (an example for this situation is a molecule suspended in vacuum between two metal leads) the observed transfer rate, flux or yield should be averaged over the initial thermal energy distribution. For higher energies the dependence on bridge length is weaker and for energies in resonance with the bridge levels it disappears. For increasing bridge length the contribution of higher energies is more important. This implies a deviation from the exponential behavior (2) that was obtained for a given $\Delta E_B$.

In the following sections we examine the dependence on bridge length arising under these two scenarios that do not involve thermal relaxation on the bridge. We find that the first possibility leads to crossover from exponential to length independence that is similar to that predicted by the thermal relaxation model. In fact we show that the experimental results of Giese and coworkers[7] on the distance dependence of electron-transfer in DNA can be fitted into this scenario. In the second case that we examine for the conduction problem we find at room temperature marked deviations from exponential behavior, however no sharp crossover behavior. It is important to emphasize that it is in principle possible to distinguish experimentally between these different effects on the bridge length dependence of electron transfer and conduction. We return to this point in the conclusion section.

## 2. Electron transfer following a sudden initial preparation of the donor state

In what follows we take $\varepsilon_D = \varepsilon_A \equiv \varepsilon_0$. We focus on the process that follows an initial population of the state $D$ (or 0). The actual process of electron transfer between donor and acceptor sites involves nuclear reorganization at these sites. This reorganization is the principal source of irreversibility in this process where other sources may arise from radiative or non-radiative decay of the donor and acceptor states or from electron capture from the bridge. Here we will consider a simpler model where decay of donor and acceptor states as well as possible decay of bridge states is incorporated by assigning complex values with negative imaginary parts to the corresponding energies $\varepsilon_0 - i\gamma_0, \varepsilon_1 - i\gamma_i, ...., \varepsilon_{N+1} - i\gamma_{N+1}$. The ($N$+2) complex eigenstates and eigenenergies of $H$ are denoted $|\psi_j\rangle$ and $E_j - i\Gamma_j$ (with real $E_j$). The time evolution that follows the initial population of the donor state is given by[8]

$$\Psi(t) = \sum_{j=0}^{N+1} \langle \psi_j | D \rangle |\psi_j\rangle e^{-(i/\hbar)(E_j - i\Gamma_j)t} \quad ; \quad \Psi(t=0) = |D\rangle \tag{3}$$

So that the time dependent probabilities that the electron is in the donor/acceptor states are

$$P_D(t) = \left| \sum_{j=0}^{N+1} R_{DD}^{(j)} e^{-(i/\hbar)(E_j - i\Gamma_j)t} \right|^2 \tag{4}$$

and

$$P_A(t) = \left| \sum_{j=0}^{N+1} R_{DA}^{(j)} e^{-(i/\hbar)(E_j - i\Gamma_j)t} \right|^2 \tag{5}$$

where

$$\begin{aligned} R_{DD}^{(j)} &= \langle \psi_j | D \rangle \langle D | \psi_j \rangle \\ R_{DA}^{(j)} &= \langle \psi_j | D \rangle \langle A | \psi_j \rangle = R_{AD}^{(j)*} \end{aligned} \tag{6}$$

In order to make contact later with the experimental work of Ref. 7 and the theoretical analysis of Ref. 4 we focus on the bridge-length dependence of the yield ratio

$$F = \frac{Y_A}{Y_D} = \frac{\gamma_A \int_0^\infty dt P_A(t)}{\gamma_D \int_0^\infty dt P_D(t)} \tag{7}$$

Note however that one could do a similar analysis in terms of the acceptor yield $Y_A = \gamma_A \int_0^\infty dt P_A(t)$. These quantities are easily calculated from Eqs. (4) and (5)

$$Y_A = \gamma_A \int_0^\infty dt P_A(t) = \gamma_A \sum_j \sum_{j'} \frac{R_{DA}^{(j)} R_{AD}^{(j')}}{(i/\hbar)(E_j - E_{j'} - i(\Gamma_j + \Gamma_{j'}))} \tag{8}$$

$$Y_D = \gamma_D \int_0^\infty dt P_D(t) = \gamma_D \sum_j \sum_{j'} \frac{R_{DD}^{(j)} R_{DD}^{(j')}}{(i/\hbar)(E_j - E_{j'} - i(\Gamma_j + \Gamma_{j'}))} \tag{9}$$

For any bridge-length N, following diagonalization of the Hamiltonian $H$ the yield $Y_A$ and the yields ratio $R$ are easily evaluated. It is also of interest to consider the case where rapid dephasing due to environmental interactions destroys all coherences in the eigenstates representation of the system's density matrix on a time scale fast relative to the electron transfer. In this case, following the preparation of the donor state, the $t=0$ density matrix can be taken diagonal and consequently

$$\rho_{inc}(t) = \sum_{j=0}^{N+1} R_{DD}^{(j)} |\psi_j\rangle\langle\psi_j| e^{-(i/\hbar)(E_j - i\Gamma_j)t} \tag{10}$$

The subscript *inc* refers to the incoherent case. This leads to the yields

$$Y_{A,inc} = \gamma_A \sum_j \frac{|R_{DA}^{(j)}|^2}{(2/\hbar)\Gamma_j} \quad ; \quad Y_{D,inc} = \gamma_D \sum_j \frac{(R_{DD}^{(j)})^2}{(2/\hbar)\Gamma_j} \tag{11}$$

which should be used in Eq. (7) to yield $F_{inc}$.

Finally, it is also of interest to examine the implication of a common approximation, the super-exchange model, to our problem. This approximation provides a good description of the transfer dynamics in the limit $|\Delta E_B| \gg |V_B|$, and is attained by replacing the $N+2$ levels description of the bridge assisted electron transfer by a two level description in which the donor and acceptor interact directly with an effective coupling determined by their coupling to the bridge and by the bridge electronic properties. The rational behind this approximation[2] is that in the weak coupling limit considered the two lowest eigenenergies (or in fact their real parts) are well separated from the rest of the spectrum and the corresponding eigenfunctions $\psi_0$ and $\psi_1$ are dominated by the donor and acceptor states, $|0>$ and $|N+1>$, see Fig. 2. Consequently, $|R_{DA}^{(0)}|, |R_{DA}^{(1)}| \gg |R_{DA}^{(j)}|$, $j \neq 0,1$ and the sums (8), (9) and (11) will be dominated by the $j=0, 1$ eigenstates. Inverting the argument, the donor and acceptor states, and their interstate dynamics can be described in the reduced representations of just these two eigenstates. The effective coupling is often identified with half the splitting between the corresponding eigenvalues. Applying this approximation to the yields defined above leads to equations identical to (8)-(9) for the coherent initial state and (11) for the incoherent initial distribution, except that now the sums over $j$ and $j'$ are limited to the two lowest eigenstates obtained from the Hamiltonian diagonalization. We denote the yields obtained in this approximation $Y_D^{su}$ and $Y_A^{su}$ (or $Y_{D,inc}^{su}$ and $Y_{A,inc}^{su}$). The corresponding ratios, Eq. (7) will be denoted $F^{su}$ and $F_{inc}^{su}$.

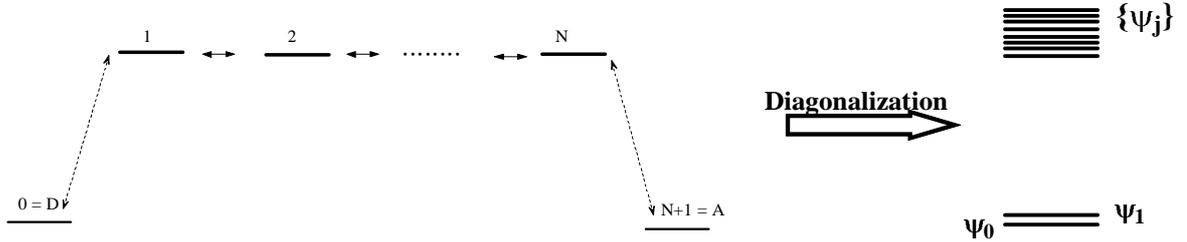

Fig. 2. Local states (left) and diagonal states (right) for the DBA model under consideration.

Figures 3 and 4 show results obtained from applying these considerations. The Hamiltonian $H_{DBA}$ is used with $\varepsilon_D = -i\gamma_D$, $\varepsilon_A = -i\gamma_A$, $\varepsilon_1 = \Delta E_B - i\gamma_1$ and $\varepsilon_l = \Delta E_B$ for $l = 2 - N$. The values of $\Delta E_B$, $V$ and $V_B$ are those used for a DNA bridge by Bixon and Jortner.[4] Figure 3 shows the results obtained by using Eqs. (8) and (9) for the full description as well as for the effective two state model. Fig. 4 shows similar results obtained for the incoherent initial condition, Eq. (11). The following observations can be made: (a) A cross over from a fast, exponential-like decrease with increasing bridge length to independence on this length is observed both for the coherent and incoherent initial distributions in the complete (*N*+2)-state calculation. (b) Such a crossover is not obtained in the super-exchange approximation where the contribution of only the two lowest eigenstates to the transfer calculation is taken into account. (c) Rapid dephasing of the initial distribution (Fig. 4) seems to have a relatively weak effect on the length dependence. The length dependence of the yield ratio is qualitatively similar in the coherent and incoherent cases; in fact it is almost the same for the parameters shown. (d) For some choice of parameters the yield ratio may oscillate as a function of bridge length (Fig. 3). This interesting interference behavior will be probably erased in realistic situations because of dephasing interactions and inhomogeneous broadening effects.

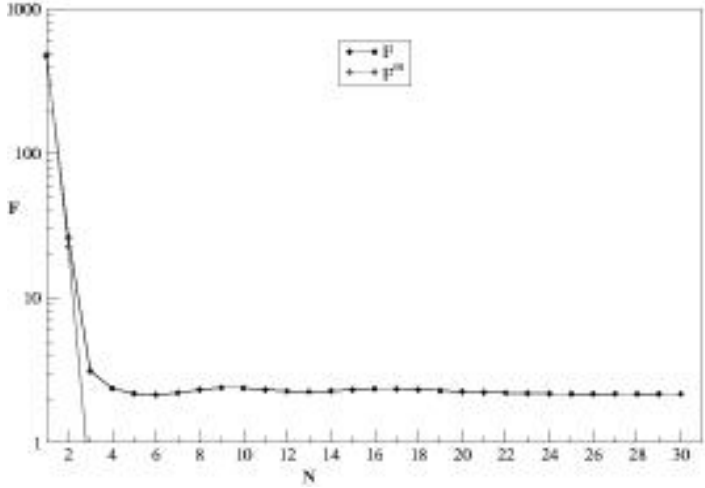

Fig. 3. Comparison of the relative yields $F$ and $F^{su}$ plotted against the number of bridge states *N* for a Hamiltonian $H_{DBA}$ with $\varepsilon_D = -i\gamma_D$, $\varepsilon_A = -i\gamma_A$, $\varepsilon_1 = \Delta E_B - i\gamma_1$ and $\varepsilon_l = \Delta E_B$ for $l = 2 - N$. The parameters chosen are: $\Delta E_B = 0.15 eV$, $V = 0.089 eV$, $V_B = 0.03 eV$, $\gamma_D = 8 \times 10^{-5} eV$, $\gamma_A = 4 \times 10^{-2} eV$ and $\gamma_1 = 1 \times 10^{-2} eV$.

1010

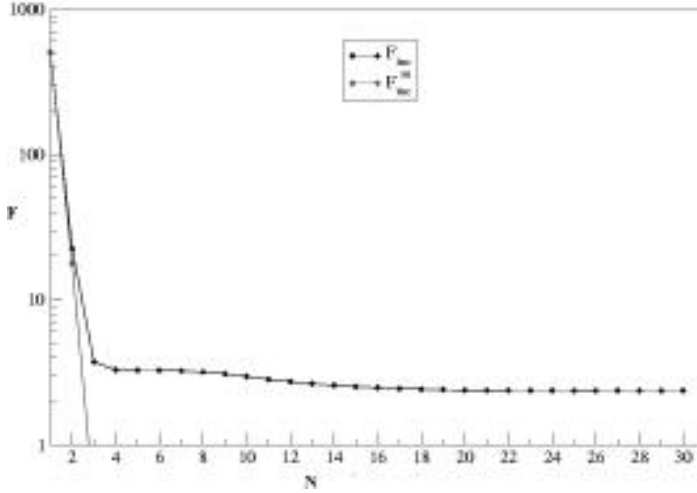

Fig. 4. Comparison of the incoherent relative yields $F_{inc}$ and $F_{inc}^{su}$ shown against the number of bridge states for the same Hamiltonian parameters as in Fig. 3.

The origin of the non-exponential behavior seen in Figs. 3 and 4 is the fact that for an initially prepared donor state is not an eigenstates of the system, the initial energy is not well defined and a small amount of population in eigenstates dominated by the bridge may be initially excited. Next we consider a different situation in which the energy is well defined.

## 3. Electron transmission in a scattering-type process

A related but different experimental setup in which electron transmission at a given energy can be monitored as a function of bridge length is a scattering-type experiment in which an electron is incident at a given energy on the DBA system from, say, the donor side. In this case the energy is well defined by the incoming state. As an example consider the experiment[9] in which electrons were injected into N-hexane films adsorbed on a polycrystaline Pt foil at energies below the bottom of the N-hexane conduction band (~0.8eV). The role of bridge states is here assumed by impurity states in the hydrocarbon band gap. The energy is determined by the incident electron beam. Another situation is depicted in Fig. 1b, where the zero temperature, zero bias conduction between the left (L) and right (R) electrodes is given by the Landauer formula

$$g = \frac{e^2}{\pi\hbar} \mathcal{T}(E_F) \tag{12}$$

where $E_F$ is the Fermi energy in the two leads, e - the electron charge and where $\mathcal{T}(E)$, a scattering theory result, is the transmission probability (summed over all initial and final channels) for an electron incident from the (say) left electrode *at energy E* to emerge on the right. The equivalent finite temperature result reads

$$g = \lim_{\Phi \to 0} \left[ \frac{e}{\pi \hbar \Phi} \int_0^\infty dE \mathcal{T}(E) \left( f(E) - f(E + e\Phi) \right) \right] = \frac{e^2}{\pi \hbar} \int_0^\infty dE \mathcal{T}(E) \frac{\partial f(E)}{\partial E} \tag{13}$$

where $f(E)$ is the Fermi distribution.

In what follows we limit ourselves for simplicity to the situation that state 1 only of the bridge is coupled to the left lead, and state $N$ only is coupled to the right lead. The transmission probability at energy $E$ is then given by[10,11]

$$\mathcal{T}(E) = |G_{N1}(E)|^2 \, \gamma_1^{(L)}(E) \gamma_N^{(R)}(E) \tag{14}$$

where $\gamma_1^{(L)}(E)$ and $\gamma_N^{(R)}(E)$ are the decay widths of levels 1 and $N$ of the bridge associated with their coupling to the corresponding left and right leads. The dependence on bridge length is obtained from the $N$ dependence of $|G_{N1}(E)|^2$. It is easy to show that to the lowest order in $|V_B/\Delta E_B|$ (where $\Delta E_B = E_B - E$) this matrix element squared is proportional to $(V_B/\Delta E_B)^{2N}$, implying the exponential decrease in bridge length given by Eq. (2). We show in the appendix that such exponential decrease is obtained as long as $|V_B| < \Delta E_B$. We conclude that the transmission at a given energy $E$ depends exponentially on the bridge length except in what is essentially a resonance situation. The reason for the different conclusion obtained in the previous section is evident. In Section 2 we have considered an initially prepared zero order ("donor") state, where by definition the energy was not well defined. The (small) initial population of eigenstates dominated by the bridge is what gave rise to the crossover from the off-resonance component that dominates the short bridge transfer to the on-resonance component that remains for long bridges. Here the energy is well defined and in off-resonance situations the dependence on bridge length is purely exponential decay. The situation is different at finite temperatures, where different injection energies contribute according to Eq. (13). Figure 5 shows the resulting behavior at room temperature. A marked deviation from the exponential behavior obtained at T=0 is seen, however in marked contrast to the situation discussed in Section 2, or to the case that involves thermal relaxation on the bridge[6] we do not observe sharp crossover from exponential to no or weak *N* dependence.

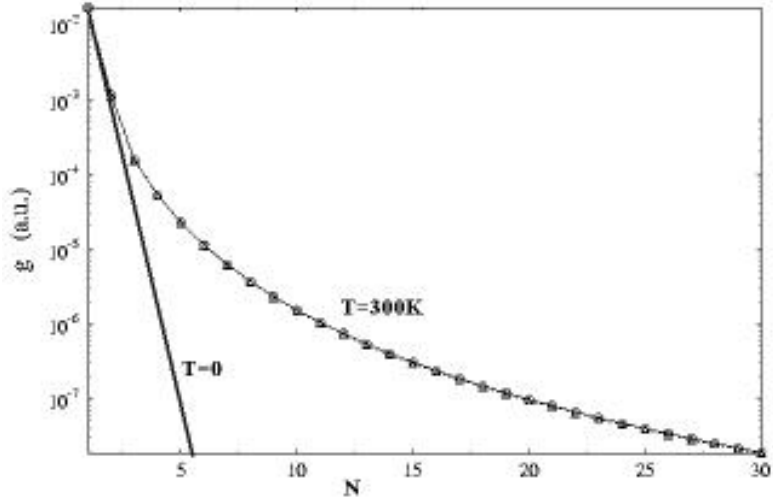

FIG 5. The zero-bias conduction obtained from Eq. (13) for the model of Fig. 1b, using the parameters: $\Delta E_B = E_B - E_F = 0.15\text{eV}$, $V_B = 0.03\text{eV}$, $\gamma_1^{(L)} = \gamma_N^{(R)} = 0.05\text{eV}$.

To end this section we note that the dependence on bridge length in these systems will obviously depend also on the applied potential bias. It will be of interest to study this issue in more detail when experimental data becomes available.

## 4. Discussion and conclusions

The dependence of bridge assisted electron transfer on the molecular bridge length and the dependence of molecular wire conduction on the wire length are obviously interesting and important attributes of these processes. In particular the crossover from exponential to very weak length dependence for increasing bridge lengths has attracted much attention recently, and was rationalized by thermal relaxation and dephasing processes in the bridge. In the present paper we have identified two other factors that affect the bridge length dependence. Both are related to the fact that the injection energy is an important parameter in this consideration.

The first factor (see Figs. 3 and 4) arises not from the physical nature of the system but from the choice of experimental setup and experimental observable. In an experiment characterized by a sudden (on the experimental timescale) preparation of the initial zero order ("donor") state the energy is not well defined, some eigenstates dominated by the bridge are also excited (viewed as a tunneling process, these are "above barrier" states), and their contribution to the transmission may dominate the electron transfer for long bridges. The relative importance of this contribution to the observed electron transfer depends of course on the system parameters. In fact, with a reasonable choice of parameters this model can successfully reproduce the experimental results of Giese et al.[7] for hole transfer in DNA duplexes consisting of Guanine donor and acceptor states separated by Adenine-Thymine bridges of varying lengths. In Ref. 7 the yield ratio of the reaction $G^+(AT)_N(GGG) \to G(AT)_N(GGG)^+$ was measured as a function of bridge length $N$ and the relative yield was found to decrease exponentially for $N < 3$ and it converges to a value of $2.5 \pm 0.5$ for $N > 3-4$. In the corresponding model given by the Hamiltonian $H_{DBA}$ (Eq. 1) the hole-donor state of $G^+$ corresponds to state $|0\rangle$, the hole-acceptor state of $(GGG)^+$ is the state $|N+1\rangle$, and the intermediate Adenine-hole $A^+$ states are represented by the bridge states $\{|l\rangle\}$. We apply the incoherent model discussed in Section 2 so that the yield ratio is modeled by $F_{inc}$, since the nature of the preparation process in Ref. 7 suggests rapid loss of electronic coherence. Using (as in figures 3 and 4) the parameters of Bixon and Jortner[4] for $\Delta E_B$, $V$ and $V_B$ it is possible to fit the experimental data of Giese et al by setting $\gamma_j = 0$ for $j = 2,...N$ and varying $\gamma_0$, $\gamma_1$ and $\gamma_{N+1}$ (see Figure 6). Note that we did not make an exhaustive search for the best fitting parameters as our main purpose here is to demonstrate the potential applicability of the suggested model. Also, it should be emphasized that this observation by no means implies that the present model is the correct interpretation of this experiment, only that it may offer a possible alternative. Different mechanisms may coexist, as discussed below.

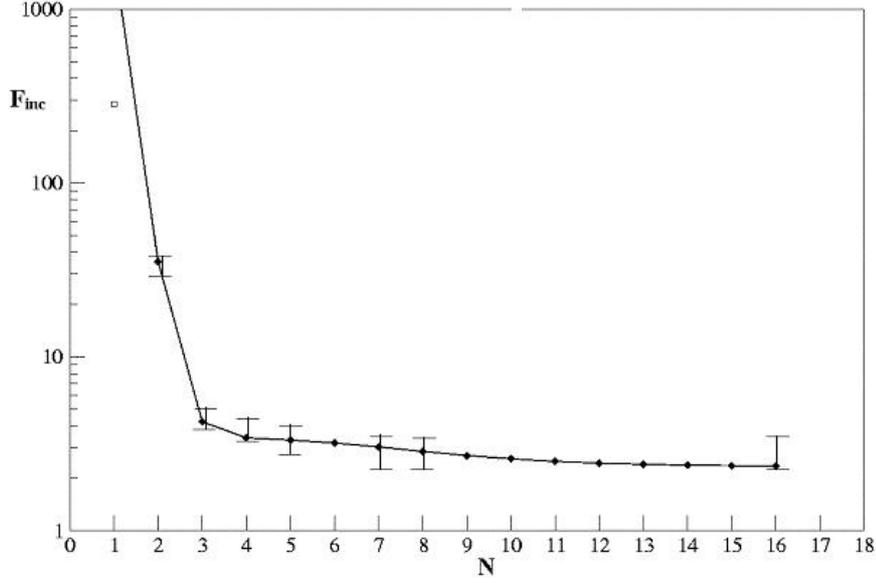

Fig. 6. The relative yield $F_{inc}$ as a function of the number of bridge states for a Hamiltonian $H_{DBA}$ with $\varepsilon_D = -i\gamma_D$, $\varepsilon_A = -i\gamma_A$, $\varepsilon_1 = \Delta E_B - i\gamma_1$ and $\varepsilon_l = \Delta E_B$ for $l = 2 - N$. The parameters chosen are: $\Delta E_B = 0.15 eV$, $V = 0.089 eV$, $V_B = 0.03 eV$, same as those used by Bixon and Jortner[4] to fit the experimental results of Ref. 7, and $\gamma_D = 2.935 \times 10^{-5} eV$, $\gamma_A = 5.87 \times 10^{-2} eV$ and $\gamma_1 = 2.2 \times 10^{-3} eV$. The error bars are the experimental results of Ref. 7. The N=1 point represented by an open square is an experimental lower bound to the actual yield ratio.

Note that the dominance of bridge-like eigenstates in the electron transfer in the long bridge limit is in fact a failure of the super-exchange model. This observation has been made for time–dependent Hamiltonians $H_{DBA}(t)$ [12,13]. For a time-dependent system where the donor and acceptor are off-resonant to the bridge at all times, the eigenstates $\{\Psi_j(t)\}$ at each time $t$ will have an eigenspectrum similar to figure 2. Two eigenstates $\Psi_0(t), \Psi_1(t)$ will be energetically separated from the rest and will have large donor and acceptor components. The time-dependent electron-transfer probability $P_A(t) = |\langle A(t)|\Psi(t)\rangle|^2$ [14] can then be written as a sum of eigenstate contributions, $P_A(t) = \left|\sum_{j=0}^{N+1} \langle A(t)|\Psi_j(t)\rangle\langle\Psi_j(t)|\Psi(t)\rangle\right|^2$, where the superexchange contribution at time $t$ is the partial sum over $\Psi_0(t), \Psi_1(t)$. For small bridge lengths the maximum amplitude of $P_A(t)$ decays exponentially with increasing bridge length because the superexchange contribution to the probability dominates the sum. At large bridge lengths superexchange is negligible and the other eigenstates mainly contribute to $P_A(t)$. In this situation the maximum amplitude of $P_A(t)$ decays very slowly with increasing bridge length. This behavior has been observed in molecular-dynamics simulations of fluctuating protein donor-bridge-acceptor systems.[13]

The first factor discussed above plays a potential role in transient experiments following an initial state preparation. The second factor is associated with initial thermal distributions and is potentially important also in steady state experiments and follows from the simple observation that in most experimental situations the initial distribution of "donor states" is not limited to a single energy even in a long time experiment where the energy can in principle be well defined. In molecular conduction the initial distribution is determined by the temperature and consequently so is the resulting dependence on wire length (Fig. 5).

It should be kept in mind that in realistic systems, effects of initial thermal distribution or of sudden preparation of the initial distribution may coexist with thermal relaxation effects. The resulting bridge length dependence will reflect the combination of these factors. Furthermore, in most experimental system changing the bridge length dependence may affect the bridge electronic structure (within the simple model considered here - the parameters $V_B$ and $\Delta E_B$). Further experimental studies of bridge length dependence of electron transfer and transmission will provide a desirable tool in elucidating the mechanism of these processes.

**Appendix**

Consider the Green's function $G_{N1}(E_F)$ in eq (14) where $E_F$ is the Fermi energy. The Green's function $G(E) = (E - H)^{-1}$ corresponds to the $N$-state bridge Hamiltonian

$$H = \begin{pmatrix} \varepsilon_1 - i\gamma_1 & V_B & 0 & \ldots & 0 \\ V_B & \varepsilon_2 & \ddots & & \vdots \\ 0 & \ddots & \ddots & \ddots & 0 \\ \vdots & & \ddots & \varepsilon_{N-1} & V_B \\ 0 & \ldots & 0 & V_B & \varepsilon_N - i\gamma_N \end{pmatrix} \quad (15)$$

where $\gamma_{1(N)}$ are the decay widths associated with the decay of the first and last bridge levels into the left and right leads. In what follows we will show that $G_{N1}(E_F)$ can be written in terms of the imaginary widths $-i\gamma_{1(N)}$ and the real Green's function matrix elements $\tilde{G}_{N-1,2}(E_F), \tilde{G}_{N-1,N-1}(E_F)$ and $\tilde{G}_{2,2}(E_F)$, where $\tilde{G}(E_F)$ is the Green's function of the Hamiltonian for a truncated bridge, without states $|1\rangle$ and $|N\rangle$

$$\tilde{H} = \begin{pmatrix} \varepsilon_2 & V_B & 0 & \ldots & 0 \\ V_B & \varepsilon_3 & \ddots & & \vdots \\ 0 & \ddots & \ddots & \ddots & 0 \\ \vdots & & \ddots & \varepsilon_{N-2} & V_B \\ 0 & \ldots & 0 & V_B & \varepsilon_{N-1} \end{pmatrix} \quad (16)$$

We will take all $\varepsilon_i$ to be the same and define $\varepsilon_i = E_F + \Delta E_B$. It is then shown below that

$$G_{N,1} = \frac{V_B}{-\Delta E_B + i\gamma_N - V_B^2 \left[ \tilde{G}_{N-1,N-1} + \delta \tilde{G} \right]} \tilde{G}_{N-1,2} \frac{V_B}{-\Delta E_B + i\gamma_1 - V_B^2 \tilde{G}_{2,2}} \quad (17)$$

where

$$\delta \tilde{G} = \tilde{G}_{N-1,2} V_B \frac{1}{-\Delta E_B + i\gamma_1 - V_B^2 \tilde{G}_{2,2}} V_B \tilde{G}_{2,N-1}. \quad (18)$$

Furthermore, we show that $\tilde{G}_{N-1,2}(E_F)$ (therefore also $\delta G$) decreases exponentially with $N$ whereas $\tilde{G}_{N-1,N-1}(E_F)$ and $\tilde{G}_{2,2}(E_F)$ are essentially $N$ independent. Eq. (17) then implies that $G_{N,1}(E_F)$ (and consequently $\mathcal{T}(E_F)$) decays exponentially with $N$.

To prove these statements define the Hamiltonian for an $m$-state system with all states coupled by the nearest neighbor interactions

$$H^{(m)} = \sum_{j=1}^{m} \varepsilon_j |j\rangle\langle j| + V_B \sum_{j=1}^{m-1} (|j\rangle\langle j+1| + hc) \quad (19)$$

(so that $\tilde{H} = H^{(N-2)}$) and the corresponding Hamiltonian in which the $m$-th state is uncoupled from the rest

$$H_0^{(m)} = \sum_{j=1}^{m} \varepsilon_j |j\rangle\langle j| + \sum_{j=1}^{m-2} (V_{j,j+1} |j\rangle\langle j+1| + hc) \qquad (20)$$

In our model $V_{j,j+1} = V_B$ for all j, but it is convenient to keep here the general notation. The relation between these two Hamiltonians is:

$$H^{(m)} = H_0^{(m)} + V^{m-1/m} \quad ; \qquad V^{m-1/m} = V_B |m-1\rangle\langle m| + hc. \qquad (21)$$

The corresponding Green's functions

$$G^{(m)} = \left[ E - H^{(m)} \right]^{-1} \text{ and } G_0^{(m)} = \left[ E - H_0^{(m)} \right]^{-1}, \qquad (22)$$

satisfy

$$G^{(m)} = G_0^{(m)} + G_0^{(m)} V^{m-1/m} G^{(m)} \qquad (23)$$

and

$$\left[ G_0^{(m)} \right]_{m,m} = \frac{1}{E - \varepsilon_m} \text{ and } \left[ G_0^{(m)} \right]_{m,j} = 0 \text{ for } j < m \qquad (24)$$

Using Eqs. (23) and (24) we can easily show that

$$\left[ G^{(m)} \right]_{m,1} = \frac{V_B}{E - \varepsilon_m} \left[ G^{(m)} \right]_{m-1,1} \qquad (25)$$

and

$$\left[ G^{(m)} \right]_{m-1,1} = \left[ G_0^{(m)} \right]_{m-1,1} + \frac{V_B}{E - \varepsilon_{m-1}} \left[ G^{(m)} \right]_{m,1}. \qquad (26)$$

This in turn lead to

$$\left[ G^{(m)} \right]_{m,1} = \frac{V_B}{E - \varepsilon_m - V_B^2 \left[ G_0^{(m)} \right]_{m-1,m-1}} \left[ G_0^{(m)} \right]_{m-1,1}. \qquad (27)$$

Note that $\left[ G_0^{(m)} \right]_{m-1,j} = \left[ G^{(m-1)} \right]_{m-1,j}$ for $j < m$. A similar approach can be used to obtain:

$$\left[ G^{(m)} \right]_{m,m} = \frac{1}{E - \varepsilon_m - V_B^2 \left[ G_0^{(m)} \right]_{m-1,m-1}} = \frac{1}{E - \varepsilon_m - V_B^2 \left[ G^{(m-1)} \right]_{m-1,m-1}}. \qquad (28)$$

Next consider the fractional change in the Green's function matrix element between the first and last states

$$\kappa_{m+1,m} \equiv - \frac{\left[ G^{(m+1)} \right]_{m+1,1} - \left[ G^m \right]_{m,1}}{\left[ G^m \right]_{m,1}} \qquad (29)$$

associated with increasing the bridge length by one element, i.e., $H^{(m)} \to H^{(m+1)}$. Using (27) we find

$$\kappa_{m+1,m} = 1 - \frac{V_{m+1,m}}{E - \varepsilon_{m+1} - V_{m+1,m}^2 \left[ G^{(m)} \right]_{m,m}} \qquad (30)$$

Taking now $E=E_B$ and $\varepsilon_m = E_F + \Delta E_B$ $V_{m+1,m} = V_B$ for all $m$ we get

$$\kappa_{m+1,m} = 1 + \frac{V_B}{\Delta E_B + V_B^2 \left[ G^{(m)} \right]_{m,m}}, \qquad (31)$$

where (c.f. Eq. (28))

$$\left[ G^{(m)} \right]_{m,m} = -\frac{1}{\Delta E_B + V_B^2 \left[ G^{(m-1)} \right]_{m-1,m-1}} \qquad (32)$$

By iterating eq (32) we conclude that as long as $V_B / \Delta E_B < 1$ $G_{m,m}^m$ decays slowly and to order $V_B / \Delta E_B$ in the denominator converges to

$$G_{m,m}^{(m)} = -\frac{1}{\Delta E_B - V_B^2 / \Delta E_B} \qquad (33)$$

with increasing $m$. Eqs. (33) and (31) then imply that $\kappa_{m+1,m}$ is approximately independent of $m$, i.e.,

$$\kappa_{m+1,m} \approx \kappa = 1 + \frac{V_B}{\Delta E_B - \dfrac{V_B^2}{\Delta E_B - V_B^2 / \Delta E_B}}. \qquad (34)$$

Consequently, from Eq. (29) we may conclude that

$$G_{m+1,1}^{(m+1)} \approx \exp[-\kappa m] G_{1,1}^1 \qquad (35)$$

This concludes the proofs of the statements made below Eq. (18). Next we prove Eqs. (17) and (18). Consider again the Hamiltonian (15) set again $\varepsilon_l = E_F + \Delta E_B$ for $l = 1...N$. In the same spirit as above we use the separation

$$H = H_0^{(N)} + V^{N-1/N}, \quad \text{with} \quad V^{N-1/N} = V_B |N-1\rangle\langle N| + hc \qquad (36)$$

where now

$$H_0^{(N)} = \tilde{H} + (\varepsilon_1 - i\gamma_1)|1\rangle\langle 1| + (\varepsilon_N - i\gamma_N)|N\rangle\langle N| + (V_B |1\rangle\langle 2| + hc) \qquad (37)$$

($\tilde{H}$ was defined by Eq. (16)). Applying Eq. (27) we get

$$G_{N,1} = \frac{V_B}{-\Delta E_B + i\gamma_N - V_B^2 G_{N-1,N-1}^{(N-1)}} G_{N-1,1}^{(N-1)}, \tag{38}$$

where $G^{(N-1)} = \left(E_F - H^{(N-1)}\right)^{-1}$ with

$$H^{(N-1)} = H_0^{(N)} - (\varepsilon_N - i\gamma_N)|N\rangle\langle N| = \tilde{H} + (\varepsilon_1 - i\gamma_1)|1\rangle\langle 1| + V^{N-2/N-1}. \tag{39}$$

where $V^{N-2/N-1} = V_B|2\rangle\langle 1| + hc$.

The matrix elements $G_{N-1,1}^{(N-1)}$ and $G_{N-1,N-1}^{(N-1)}$ are obtained by applying Eq. (23) and its alternative form $G^{(m)} = G_0^{(m)} + G^{(m)} V^{m-1/m} G_0^{(m)}$. The final result is

$$G_{N-1,1}^{(N-1)} = \tilde{G}_{N-1,2} \frac{V_B}{-\Delta E_B + i\gamma_1 - V_B^2 \tilde{G}_{2,2}} \tag{40}$$

and

$$G_{N-1,N-1}^{(N-1)} = \tilde{G}_{N-1,N-1} + \tilde{G}_{N-1,2} \frac{V_B^2}{-\Delta E_B + i\gamma_1 - V_B^2 \tilde{G}_{2,2}} \tilde{G}_{2,N-1} \tag{41}$$

Substituting eqs (40) and (41) into (38) leads to Eqs. (17) and (18).

**Acknowledgements.** This research was supported by the US-Israel Binational Science Foundation, By the Israel Ministry of Science and by Israel Science Foundation, by the University of Cyprus and by Institute of Chemical Physics, TAU. We thank Dr. M. Galperin for his help in producing Fig. 5 and Professors M. Bixon and J. Jortner for helpful discussions.

# References

1. See, e.g., A. Nitzan, *Ann. Rev. Phys. Chem.* 52, 681- 750 (2001).

2. See, e.g., S. S. Skourtis and D. N. Beratan, *Adv. Chem. Phys.* 106, 377-452 (1999).

3. B. Giese, *Accounts of Chemical Research* 33, 631-636 (2000).

4. M. Bixon and J. Jortner, *Chemical Physics* 281, 393-408 (2002).

5. A. K. Felts, W. T. Pollard, and R. A. Friesner, *J. Phys. Chem.* 99, 2929-2940 (1995).

6. D. Segal, A. Nitzan, W. B. Davis, M. R. Wasilewski, and M. A. Ratner, *J. Phys. Chem. B* 104, 3817 (2000).

7. B. Giese, J. Amaudrut, A.-K. Kohler, M. Spormann, and S. Wessely, *Nature* 412, 318-20 (2001).

8. Note that $\psi_0$, the lowest energy eigenstate (see Fig. 2), should not be confused with $|0\rangle$, the donor state.

9. L. G. Caron, G. Perluzzo, G. Bader, and L. Sanche, *Phys. Rev. B* 33, 3027-3038 (1986).

10. V. Mujica, M. Kemp, and M. A. Ratner, *J. Chem. Phys.* 101, 6849-6855 (1994).

11. V. Mujica, M. Kemp, and M. A. Ratner, *J. Chem. Phys.* 101, 6856-6864 (1994).

12. Q. Xie, G. Archontis, and S. S. Skourtis, *Chemical Physics Letters* 312, 237-46 (1999).

13. S. S. Skourtis, Q. Xie, and G. Archontis, *J. Chem. Phys.* 115, 9444 (2001).

14. $|\Psi(t)\rangle$ is the solution to $i\hbar = d|\Psi(t)\rangle/dt = H_{DBA}(t)|\Psi(t)\rangle$ where $|\Psi(t=0)\rangle = |D(0)\rangle$.